\documentclass[%
 reprint,
 amsmath,amssymb,
 aps,
]{revtex4-2}

\usepackage{graphicx}
\usepackage{dcolumn}
\usepackage{bm}
\usepackage{derivative}
\usepackage{gensymb}

\begin{document}

\preprint{APS/123-QED}

\title{Measuring electron density and electromagnetic fields in low-density laser-driven plasmas using proton radiography}

\author{Joshua Ting Yu Chu}
\affiliation{%
Department of Physics, University of Oxford, OX1 3PU, UK
}%

\author{Petros Tzeferacos}%
\affiliation{%
Department of Physics and Astronomy, University of Rochester, Rochester, NY, 14627, USA}
\affiliation{%
Department of Physics, University of Oxford, OX1 3PU, UK
}%

\author{Gianluca Gregori}%
\author{Archie F. A. Bott}%
\affiliation{%
Department of Physics, University of Oxford, OX1 3PU, UK
}%

\date{\today}
\nocite{Chu2026}

\begin{abstract}
In this paper, we describe a novel technique for inferring electron density in laser plasmas from proton-radiography data. Proton radiography is primarily a diagnostic for imaging electromagnetic field structures in such plasmas. In the presence of strong field gradients, `caustic' features arise and are broadened by the process of collisional scattering between the imaging protons and the plasma. This effect has a well defined density dependence, allowing for a robust measurement of electron number density.
We validate the method on data from the turbulent-dynamo experiments performed at the Omega laser facility (with electron densities in the range of ${\sim}10^{20}\,\mathrm{cm^{-3}}$), achieving good agreement with values obtained from Thomson scattering, x-ray imaging and simulations. This research enables one of the most robust techniques for measuring electromagnetic fields on laser-plasma experiments to simultaneously provide density measurements along the same line of sight, improving our ability to diagnose such plasmas.    
\end{abstract}

\maketitle


\section{\label{sec:level1}Introduction}
In the last few decades, laboratory experiments involving high-energy-density (HED) plasmas have become a dynamic area of plasma research, driven by the pursuit of inertial confinement fusion (ICF). With the advent of powerful laser and pulsed-power systems, we can achieve the high pressures ($> 1$ Mbar) and temperatures ($> 10$ eV) characteristic of HED plasmas, probing various hydrodynamic instabilities, particle dynamics, and recreating scaled astrophysical phenomena in the laboratory \cite{Ryutov1999, Gregori2015,Lebedev2019, Takabe2021, Blackman2022}. This technological advancement brings an accompanying diagnostic challenge, given the characteristic sub-micro-second timescales and sub-cubic-centimeter volumes of laser-driven HED plasmas: how do we measure their basic properties? 

Accurate characterization of the plasma electron number density is crucial for understanding and control of plasma behavior. There are a host of existing diagnostics which provide coverage for a range of density regimes. For low-density plasmas, common techniques include Langmuir probes, microwave interferometry and optical emission spectroscopy \cite{Manos1991, Meulenbroeks1994, Chai2019}. In the intermediate range ($10^{18}$-$10^{21}\,\mathrm{cm^{-3}}$) that is of our particular interest, laser-based diagnostics such as Thomson scattering \cite{Glenzer2009, Suttle2021} and interferometry \cite{Swadling2013, Hare2019} provide non-perturbative measurements. Towards the higher densities of the warm dense matter (WDM) regime, using conventional optical and probe diagnostics becomes increasingly difficult; penetrating X-ray techniques such as X-ray Thomson scattering (XRTS) \cite{Glenzer2009, Kritcher2008} and X-ray-absorption near-edge spectroscopy (XANES) \cite{BenuzziMounaix2011} are commonly employed. Nevertheless, alternative methods of measuring this plasma parameter remain important, as they can be used to validate, or overcome the limitations of existing optical diagnostics. For example, Thomson scattering cannot be used for low-temperature or high-density plasmas, as laser heating from the probe beam can significantly perturb the plasma \cite{Murphy2004}; laser interferometry can suffer from fringe jump errors when density is too high, or phase variations due to mechanical vibrations \cite{deGroot1995}. From a practical standpoint, these diagnostics might be difficult to field if there are issues related to the setup geometry, alignment, or even unforeseen circumstances such as equipment failure. 

Another important aspect of laboratory HED plasmas is the dynamics of the electromagnetic (EM) fields. In ICF, magnetic fields can be used to confine and heat the plasma \cite{Slutz2010, Chang2011, Moody2022}, increasing potential yield. EM fields also play a crucial role in many fundamental processes of interest, including shocks, filamentation instabilities, magnetic reconnection and turbulence \cite{Fiuza2012, Huntington2015, Parker1957, Tzeferacos2018}.
One diagnostic for EM fields is proton radiography (PRAD), which allows for detailed characterization of field structures. PRAD is widely implemented in HED experiments \cite{Borghesi2001, MacKinnon2004, MacKinnon2006, Li2006, Bott2021a, Bott2021b} because it offers a number of advantages compared to other diagnostics. It is non-perturbative with good spatial and temporal resolution, and is currently the only viable option for imaging both magnetic and electric fields concurrently \cite{Schaeffer2023}. Typically, a spatially uniform source of energetic protons is directed into the experimental plasma, where they interact via Lorentz forces associated with electromagnetic fields and are deflected, forming a non-uniform flux image at the detector further downstream. 
This flux image can then be analyzed to retrieve information on EM field structures, either through forward propagation \cite{Borghesi2002, Huntington2015, Romagnani2005} or reconstructed analytically \cite{Kugland2012, Graziani2017, Bott2017,davies2023}. This paper will focus on experiments \cite{Bott2022, Bott2021b} involving stochastic magnetic fields, for which analytical reconstruction is better suited. In practice, complications can arise and reconstruction is not always a tractable problem; if deflections are too large, the trajectories of the imaging protons can cross before reaching the detector, leading to intense flux structures known as caustics. 

Although transverse Lorentz scattering is the dominant process for most plasmas of interest, other beam-plasma interactions can also occur, opening up the possibility of using backlighting protons as a probe for density. Direct extraction of density from PRAD data has been done in the past on ICF experiments~\cite{MacKinnon2004, Li2008, Rygg2008, Li2010, Volpe2011}, by measuring the energy loss of the imaging protons caused by electric fields associated with high density and pressure gradients. Asides from being contingent on the existence of these strong gradients, the effects of competing processes such as nuclear and Coulomb scattering have to be accounted for (or discounted) through the use of simulations. Similar methods have also successfully measured high density gradients in shock compression experiments \cite{MacKinnon2004, LePape2006, Rigg2008, Ravisio2010}. These experiments were also in a high density regime ($n_e \gtrsim 10^{22}\,\mathrm{cm^{-3}}$) and did not measure EM fields concurrently. 

 In this paper, we build upon ideas first considered in Bott et al.~\cite{Bott2021a}, which attempted to measure areal density concurrently to magnetic fields in experiments investigating supersonic plasma turbulence by considering collisional scattering of protons used for radiography. This manifests as a broadening of the aforementioned caustic structures in proton-flux images. The known density dependency of this effect can be exploited to produce an estimate for the free electron density of the plasma. On this basis, we present a general technique for using proton radiography as a density diagnostic in addition to its main application as a probe for EM fields. 
 
The paper is structured as follows. Section~\ref{sec:theory} describes the theory behind PRAD and collisional broadening. In~\ref{sec:theory:recon}, we briefly outline the method for reconstructing stochastic magnetic fields. Next, we discuss formation of `caustic' structures in proton-flux images and methods of identifying these structures in section~\ref{sec:theory:caustics}. In section~\ref{sec:theory:coll_scatter}, we derive the scattering angle of the imaging protons undergoing small angle Coulomb collisions. Section \ref{sec:theory:caustics_ps} describes the effect of collisional scattering on the characteristic spectrum of caustic features in the proton flux images. Section~\ref{sec:theory:exp_lim} introduces caveats to applying this relation as a density diagnostic, associated with experimental limitations. Section~\ref{sec:valmethod} will provide validation of the method, using data from two experiments and magnetohydrodynamic (MHD) simulations using the code FLASH~\cite{Fryxell2000,Tzeferacos2015}. Section~\ref{sec:valmethod:sims} validates the method on synthetic proton radiography data~\cite{Tzeferacos2017}, produced by simulating the transport of imaging protons through a simple plasma with prescribed magnetic fields and density. In section~\ref{sec:valmethod:exp} the experiments are discussed; section~\ref{sec:valmethod:exp:setup} shows the relevant  setups, section~\ref{sec:valmethod:exp:caustics} demonstrates evidence for the presence of caustic structures in the flux images obtained experimentally, and section~\ref{sec:valmethod:exp:data} goes through the method for obtaining a density measurement and compares the result with other experimental diagnostics and simulations.

\section{Theory} \label{sec:theory}
\subsection{Reconstruction of stochastic fields} \label{sec:theory:recon}
We begin by illustrating the basic concepts underpinning the analytical reconstruction of stochastic magnetic fields. Firstly, given multi-MeV initial proton energies, it can be shown that the Lorentz deflections are dominated by the magnetic field component for the magnetic field strengths typically realized in laser-plasma experiments~\cite{Bott2017, Schaeffer2023}. The initial, quasi-uniform distribution of the proton flux is restructured by the magnetic fields in the plasma, leading to a final, non-uniform proton beam arriving at the detector. For arbitrary field configurations and imaging setups, the mapping between initial and final flux distributions is a very complicated problem, so the following assumptions are made. The imaging setup is typically configured in the paraxial limit, such that the characteristic length scale $\ell_{\perp}$ of the plasma in the direction perpendicular to the proton-beam trajectory satisfies $\ell_{\perp}\ll r_i$, the distance from the proton source to the plasma. We also assume a point-projection geometry ($\ell_z\ll r_s$), where the path length of the protons through the plasma $l_z$ is much less than the distance $r_s$ from the plasma to the detector. Using these assumptions, we arrive at the plasma-detector mapping \cite{Kugland2012, Bott2017}, which relates the final perpendicular position of an imaging proton $\bm{x_{\perp}^{(s)}}$ to its initial position $\bm{x}_{\perp 0}$:
\begin{equation}
    \bm{x}_{\perp}^{(s)}(\bm{x}_{\perp 0}) \approx \mathcal{M}\bm{x}_{\perp 0}+\frac{r_s}{v_0}\bm{w}(\bm{x}_{\perp 0}),
\end{equation}  
where $\mathcal{M}\approx (r_s+r_i)/{r_i}$ is the magnification of the imaging setup, $v_0$ is the initial proton-beam velocity, and $\bm{w}(\bm{x}_{\perp 0})$ is the perpendicular velocity deflection with initial position $\bm{x}_{\perp 0}$, proportional to the line-integrated magnetic field along the proton trajectory. Then, by conservation of proton flux, we arrive at the image-flux relation \cite{Kugland2012, Bott2017}, which describes the relation between the image-flux distribution and the path-integrated magnetic field:
\begin{equation}
    \Psi(\bm{x}_{\perp}^{(s)}(\bm{x}_{\perp 0})) = \sum_{\bm{x}_{\perp}^{(s)}=\bm{x}_{\perp}^{(s)}(\bm{x}_{\perp 0})} \frac{\Psi_0}{|\text{Det}(\bm{\nabla}_{\perp 0}[\bm{x}_{\perp}^{(s)}(\bm{x}_{\perp 0})])|},
\end{equation}
where $\Psi_0$ is the initial flux distribution, $\bm{\nabla}_{\perp 0}\equiv \partial/\partial \bm{x}_{\perp 0}$ is a gradient operator, and the summation indicates that the flux at $\bm{x}_{\perp}^{(s)}$ can have contributions from different $\bm{x}_{\perp 0}$.

The properties of the image-flux relation and the characteristics of image-flux features are neatly captured by a dimensionless parameter $\mu$, the contrast parameter; $\mu$ effectively compares the magnitude of deflections by the magnetic fields to the (magnified) size of the field structures, and is defined as
\begin{equation}
    \mu \equiv \frac{r_s\delta\theta}{\mathcal{M} \ell_B}, \label{constrastparam}
\end{equation}
where $\ell_B$ is the magnetic coherence length and $\delta\theta$ is the typical deflection angle of an imaging proton. For a stochastic magnetic field, it can be shown that, in the small deflection limit and under the assumption $l_z \ll l_B$, the root-mean-square (RMS) value of $\delta\theta$ is given by
\begin{equation}
    \delta\theta_{\rm rms} = \frac{eB_{\text{rms}}}{m_pcv_0}\sqrt{l_zl_B}, \label{defl_ang}
\end{equation}
where $B_{\text{rms}}$ is the RMS magnetic-field strength and $m_p$ the proton mass. This relation has been investigated experimentally and verified \cite{Chen2020}. The magnitude of $\mu$ systematically categorizes image-flux features that arise from arbitrary magnetic-field structures into different contrast regimes: linear, non-linear injective, caustic, and diffusive regimes. In this work, we will primarily consider the caustic regime, which arises when $\mu$ is greater than some critical value $\mu_c$ such that the plasma-image mapping is no longer injective.

\subsection{Caustics} \label{sec:theory:caustics}
When $\mu \gtrsim \mu_c \sim 1$, gradients in the perpendicular deflection field are sufficiently large that regions of the proton beam can self-intersect before reaching the detector. As a result, the plasma-image mapping can become multi-valued and proton-fluence inhomogeneities known as caustics can appear. For the purposes of reconstruction analysis, this poses a challenging problem, although several computational solutions, albeit non-general, have been explored~\cite{Kasim2017, Chen2017, Levesque2021}. The multi-valuedness means that for an arbitrary stochastic field, solutions obtained via the image-flux relation are no longer unique. This means that the morphology, intensity and position of caustic structures might not be representative of the field structures responsible for their formation. In addition, it can be shown analytically that the reconstructed deflection field always provides a lower bound of the RMS deflection field \cite{Bott2017}, thus reconstruction will underpredict $B_\mathrm{rms}$ in the presence of caustics.

Caustic structures are typically identifiable from proton images as narrow curves of abnormally high proton flux, and typically are formed in pairs~\cite{Kugland2012}. Generally, for proton images containing caustics, the width of image-flux structures increases with decreasing proton energy. A more robust approach of identification requires multiple proton energies sampling the same fields, i.e., the transit time differences between protons of different energies are small compared to the dynamical timescales of the magnetic fields. For example, a D\textsuperscript{3}He fusion capsule located 1 cm away from the plasma producing 3.3 and 15.0 MeV protons leads to a time difference of ${\sim}200\,\mathrm{ps}$, whereas dynamical timescales in long-pulse laser-driven setups are on the order of nanoseconds. 
Following (\ref{constrastparam}) and (\ref{defl_ang}), $\mu$ is inversely proportional to $v_0$ and the initial proton energy, and so we can consider the regime where $\mu > \mu_c$ for imaging protons with lower initial energy but $\mu < \mu_c$ for the higher energy protons. Given the absence of caustics in the high-energy flux image, we can designate the reconstructed field as the `true' path-integrated field which is similarly experienced by the low-energy protons. This reconstructed field can be used to predict the proton flux image for the slower protons (with an artificial blurring associated with the finite source size applied as outlined in section~\ref{sec:theory:exp_lim}) and should be consistent with the measured flux image. On the other hand, predicting the flux image of the higher energy protons with the reconstructed field from the lower energy protons will not produce image-flux structures with the correct morphology or width, indicating the presence of caustics in the lower-energy flux image.

\subsection{Collisional scattering} \label{sec:theory:coll_scatter}
As the imaging protons traverses the plasma with electron number density $n_e$, diffusion in the perpendicular direction will accumulate as a series of small-angle Coulomb collisions. Given the large number of scattering events, a random-walk treatment can be applied in weakly coupled plasmas. This results in a characteristic spread in perpendicular velocity $\Delta v_{\perp}$ given by
\begin{equation}
    \odv{\Delta v_{\perp}^2}{t}=(\nu_{\perp}^{p\vert i}+\nu_{\perp}^{p\vert e})v(t)^2, \label{eq:rms_deflection}
\end{equation}
where $v(t)$ is the speed of the imaging protons at time $t$ after the start of interaction with the plasma, $\nu_{\perp}^{p\vert i}=\sum_{s}\nu_{\perp}^{p\vert i_s}$ is the total perpendicular diffusion rate due to Coulomb collisions with each ion species $i_s$ and $\nu_{\perp}^{p\vert e}$ is the perpendicular diffusion rate due to collisions with electrons~\cite{Huba2013}. Assuming $v_0$ is much greater than the plasma thermal velocity, we use the diffusion rates for fast test particles:
\begin{equation}
    \nu_{\perp}^{p\vert i_s} = 1.8 \times 10^{-7} \epsilon_p^{-3/2} \langle Z_{i_s} \rangle^2 n_{i_s} \lambda_{p \vert i_s},
\end{equation}
\begin{equation}
    \nu_{\perp}^{p\vert e} = 1.8 \times 10^{-7} \epsilon_p^{-3/2} n_e \lambda_{p \vert e},
\end{equation}
where $\epsilon_p$ is the proton energy in eV, $\langle Z_{i_s} \rangle$ the mean charge of ion species $i_s$, $n_{i_s}$ and $n_e$ the ion and electron number densities, and $\lambda_{p \vert i_s}$ and $\lambda_{p \vert e}$ the proton-ion and proton-electron Coulomb logarithms, respectively~\cite{Huba2013}. If Coulomb collisions are sufficiently weak such that $v(t)\simeq v_0$ over the interaction length $\ell_n$, it follows from (\ref{eq:rms_deflection}) that 
\begin{equation}
    \Delta v_{\perp} \simeq \sqrt{(\nu_{\perp}^{p\vert i}+\nu_{\perp}^{p\vert e})\ell_{n}v_0},
\end{equation}
and therefore the characteristic scattering angle $\Delta\theta_{C}$ associated with collisional scattering is given by
\begin{equation}\label{eq:thetaC}
    \Delta \theta_C \simeq \frac{\Delta v_{\perp}}{v_0} \simeq \sqrt{\frac{(\nu_{\perp}^{p\vert i}+\nu_{\perp}^{p\vert e})\ell_{n}}{v_0}}. 
\end{equation}

\subsection{Effect of collisional scattering on the power spectrum of caustics} \label{sec:theory:caustics_ps}
Another key feature of caustics is the power spectrum of the proton-flux distribution in which they appear. In the presence of caustic structures, the power spectrum follows a characteristic $k^{-2}$ power law~\cite{Bott2017} --- see the spectra in Fig. \ref{fig:19_sim} and \ref{fig:21_sim}E, shown in red. Assuming that $\Delta \theta_C \ll 1$, the effect of collisional scattering on the proton flux distribution can be found from kinetic theory~\cite{Bott2017}. The result is a convolution of the unperturbed flux distribution with a Gaussian point-spread function, which has a full-half-width-maximum (FWHM) of $a_\text{res}$ given by:
\begin{equation}
    a_\text{res} \approx 2.2r_s\Delta \theta_C=2.2r_s\sqrt{\frac{(\nu_{\perp}^{p\vert i}+\nu_{\perp}^{p\vert e})\ell_{n}}{v_0}}.
\end{equation}
Equivalently, this corresponds to a wavenumber cutoff of $k_\text{res} \approx 4.4 \mathcal{M}/a_\text{res}$. We assume that applying this effect to the one-dimensional relative flux spectrum (in the absence of caustic structures) $\hat{E}_{\delta \Psi}$ will result in a measured spectrum $E_{\delta \Psi}$ of
\begin{equation}\label{eq:fit}
    E_{\delta \Psi}(k) = \hat{E}_{\delta \Psi}(k)\exp\left(-\frac{k^2}{k^2_\text{res}}\right) + E_\text{noise},
\end{equation}
where $E_\text{noise}$ represents the noise floor of the spectrum, which dominates at high $k$, producing a flat tail in the spectrum (see Fig. \ref{fig:densityprocess}E and F) .

\subsection{Experimental limitations} \label{sec:theory:exp_lim}
While sections~\ref{sec:theory:coll_scatter} and~\ref{sec:theory:caustics_ps} provide a theoretical treatment relating $n_e$ to $E_{\delta \Psi}$, there are additional constraints imposed by the experimental implementation of the proton-imaging diagnostic. We outline two relevant experimental complications: Poisson noise and a finite proton source size. 
\subsubsection{Poisson noise}
The imaging protons impinging on the detector can be treated as a finite number of discrete events, and thus obey Poisson statistics. Therefore, given a mean flux per pixel $N_{\Psi}$, the standard deviation of fluctuations is $\sqrt{N_{\Psi}}$, while the variance of the integrated power spectrum of is $1/N_{\Psi}$. This leads to an expression for the Poisson noise floor, which is uniform over $k$ due to the uncorrelated nature of the noise:
\begin{equation}
    E_\text{noise} = \left(\frac{\delta x}{2\pi}\right)\frac{1}{N_{\Psi}}, 
\end{equation}
where $\delta x$ is the pixel side length. This imposes a requirement that the mean proton flux and signal strength on the detector is sufficiently high such that the Poisson contribution to the spectrum is small.

\subsubsection{Finite source size}
The plasma-image mapping (1) assumes the imaging protons originate from a point source, while real experimental sources using target normal sheath acceleration (TNSA) and D\textsuperscript{3}He fusion implosions have some finite source size $d_\mathrm{s}\simeq 40-50\,\mathrm{\mu m}$ for D\textsuperscript{3}He fusion implosions \cite{Manuel2012, Seguin2004, Li2006} and $d_\mathrm{s}\simeq 10\,\mathrm{\mu m}$ for TNSA \cite{Schaeffer2023}. Although this is a valid estimate as $d_\mathrm{s} \ll r_i$, this places a lower limit on the smallest resolvable field structure, and notably introduces a smearing effect to the measured proton flux image. 

It can be shown using kinetic theory \cite{Bott2017} that the measured image-flux distribution $\tilde{\Psi}(\bm{x}_{\perp}^{(s)})$ is the result of convolving the distribution $\Psi(\bm{x}_{\perp}^{(s)})$ generated from a point source with a point-spread function $S$:
\begin{equation}
    \tilde{\Psi}(\bm{x_{\perp}^{(s)}}) = \int \text{d}^2\tilde{\bm{x}}_{\perp}^{(s)}\,\Psi(\tilde{\bm{x}}_{\perp}^{(s)}) S(\bm{x}_{\perp}^{(s)}-\tilde{\bm{x}}_{\perp}^{(s)}), 
\end{equation}
and that $S$ is given by:
\begin{equation}
    S(\bm{x}_{\perp}^{(s)}-\tilde{\bm{x}}_{\perp}^{(s)})=\frac{v^2}{r_s^2}P\left(\frac{\bm{x}_{\perp}^{(s)}-\tilde{\bm{x}}_{\perp}^{(s)}}{r_s}v\right),
\end{equation}
where $P(\delta\bm{v}_{\perp0})$ is the distribution function of perpendicular velocities. For implosion capsules, $S$ is experimentally observed to be a Gaussian with a FWHM $a_\mathrm{s}\simeq d_\mathrm{s}\mathcal{M}$. As this directly competes with the collisional scattering effect, a finite source size introduces a lower limit for the measurable density: $n_e$ has to be sufficiently high such that the corresponding $a_\text{res} \gg a_\mathrm{s}$.

\subsubsection{Finite energy bandwidth}
Experimental proton sources are not only finite spatially and temporally, but also have a finite energy bandwidth. For D\textsuperscript{3}He fusion implosions, the energy spread is up to ${\sim}3 \%$; for TNSA protons which have a broadband energy spectrum, the effective bandwidth comes from the detector side: each layer of the radiochromic film (RCF) stack has protons with energy around the Bragg peak energy deposited, with a spread of ${\sim}0.2\,\mathrm{MeV}$. Given the $1/v_0$ dependence in (4), this introduces another blurring effect to the proton-flux image. Assuming a spread in velocities $\Delta v$ due to finite energy bandwidth, the spread in the characteristic deflection angle is given by
\begin{equation}
    \Delta\delta\theta_E \sim \delta\theta\frac{\Delta v}{v_0},
\end{equation}
where $\Delta v/v_0\simeq \Delta E/2E_0$. Thus the associated blurring size is $a_E\sim r_s\Delta\delta\theta_E$. To compare this effect to the blurring associated with the finite source size, we evaluate 
\begin{equation}
    \frac{a_E}{a_\mathrm{s}} \simeq \frac{\mu\ell_B\Delta v/v_0}{d_s}.
\end{equation}
For a D\textsuperscript{3}He source with $\mu\sim1$ and $\ell_B=100\,\mathrm{\mu m}$, $a_E/a_\mathrm{s}\sim0.03$. Thus the contribution from the finite energy bandwidth of imaging protons is negligible compared to the effects from collisional scattering and the finite source size for typical experimental parameters.  

\section{Validation of Method} \label{sec:valmethod}
This section will demonstrate the validity of the prescription outlined in section~\ref{sec:theory}. We perform this validation using data obtained from the OMEGA laser facility and FLASH MHD simulations. Field reconstruction was performed using the PROBLEM code \cite{Bott2017}, a non-linear field-reconstruction algorithm. 

\begin{figure}[h]
    \centering
    \includegraphics[width=\linewidth]{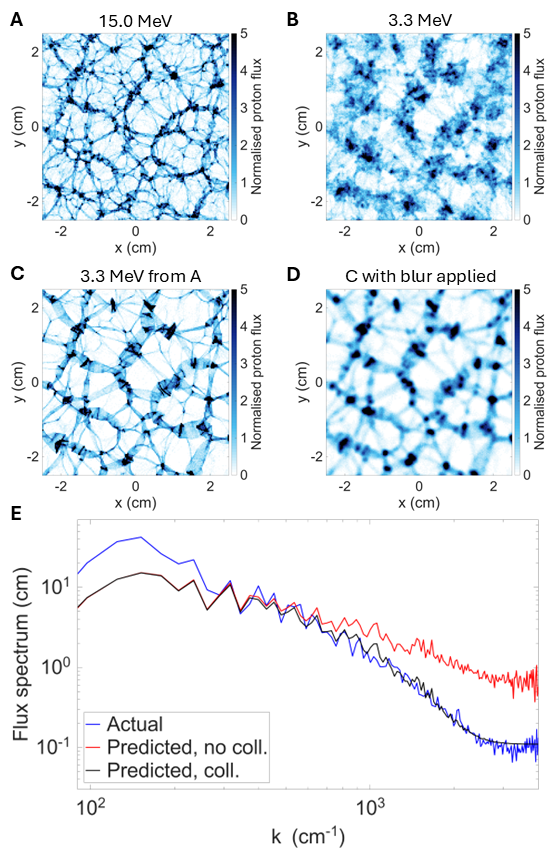}
    \caption{Simulation results for a electron number density $n_e=10^{19}\,\mathrm{cm^{-3}}$. (A) and (B) show synthetic PRAD data produced by simulating the interaction between 3.3 and 15.0 MeV proton beams and a cube of magnetized plasma. (C) shows a predicted $3.3\,\mathrm{MeV}$ PRAD image using the reconstructed deflection field obtained from running the PROBLEM code on (A). This is what (B) would be expected to look like in the absence of collisional broadening of caustic features. (D) shows (C) with a Gaussian blur applied, with the scale length obtained from the least-squares fit in (E); this image is to be compared with (B) and shows similar sizes in the blurred features. (E) shows the flux power spectra of (A) in blue, (C) in red representing the predicted unbroadened spectrum, and the least-squares best fit for the broadened spectrum shown as the black line.}
    \label{fig:19_sim}
\end{figure}

\subsection{Validation on simulated data} \label{sec:valmethod:sims}
\begin{figure}[h]
    \centering
    \includegraphics[width=\linewidth]{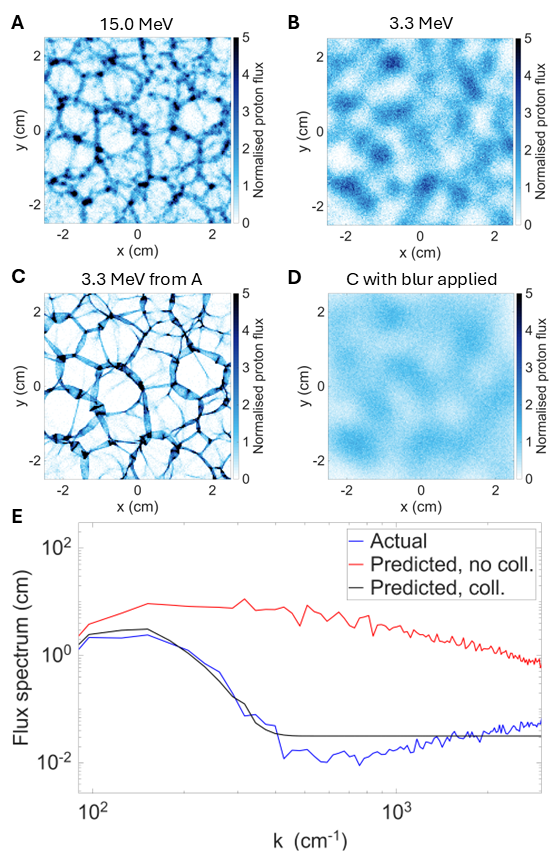}
    \caption{Simulation results for a electron number density $n_e=10^{21}\,\mathrm{cm^{-3}}$. (A) and (B) show synthetic PRAD data produced by simulating the interaction between 3.3 and 15.0 MeV proton beams and a cube of magnetized plasma. (C) shows a predicted 3.3 MeV PRAD image using the reconstructed deflection field obtained from running the PROBLEM code on (A). This is what (B) would be expected to look like in the absence of collisional broadening of caustic features. (D) shows (C) with a Gaussian blur applied, with the scale length obtained from the least-squares fit in (E); this image is to be compared with (B) and shows similar sizes in the blurred features. (E) shows the flux power spectra of (A) in blue, (C) in red representing the predicted unbroadened spectrum, and the least-squares best fit for the broadened spectrum shown as the black line.}
    \label{fig:21_sim}
\end{figure}

We first test the method on synthetic PRAD data, which allows for a robust validation of the method in measuring a known density, in the absence of blurring associated with the finite source size. To accomplish this, we developed a simulation that couples a Boris pusher to small-angle collisional scattering. Beams of $10^7$ protons originating from a spherically expanding point source are initialized with energies of $15.0\,\mathrm{MeV}$ and $3.3\,\mathrm{MeV}$, corresponding to the protons produced from nuclear fusion reactions of D\textsuperscript{3}He. After propagating freely for 1 cm, the protons traverse a cube of uniform-density carbon plasma, 2 mm across, in which they are deflected by a static, stochastic Gaussian magnetic field. The field has a three-dimensional Kolmogorov-like spectrum which follows a $k^{-11/3}$ power law, with $B_\mathrm{rms}=100\,\mathrm{kG}$ and a single characteristic scale length of $\ell_{\rm B} = 0.5\,\mathrm{mm}$. After each Boris push step, collisional scattering is also implemented as a continuous random walk: a deflection is randomly chosen from a normal distribution with a magnitude given by (\ref{eq:thetaC}), under the realistic plasma conditions of $T_e =100\,\mathrm{eV}$ and full ionization ($\bar{Z}=6$). Subsequently, the protons freely propagate to a virtual detector located 27 cm downstream of the plasma, with the beam subtending an angle that fills the $6\,\mathrm{cm}$ wide square detector; the setup geometry results in a magnification of $\mathcal{M}=28$. Hits on the detector are binned at a similar resolution to typical experimental values and Poisson noise is added to produce the synthetic PRAD data show in (A) and (B) of Fig. \ref{fig:19_sim} and \ref{fig:21_sim}. 

First, the magnetic-field reconstruction is performed on the synthetic PRAD images using the PROBLEM code. The $15.0\,\mathrm{MeV}$ data reproduces the correct magnetic field strength of $B_\mathrm{rms}=100\,\mathrm{kG}$; the field strength is suppressed by a factor of approximately two in the $3.3\,\mathrm{MeV}$ energy case, indicating the presence of caustics. Next, the reconstructed deflection field from $15.0\,\mathrm{MeV}$ data is assumed to be the same field that the $3.3\,\mathrm{MeV}$ protons encounter and can be used to predict the $3.3\,\mathrm{MeV}$ proton image in the absence of caustic broadening: 3.3 MeV protons are initialized in the plane of the deflection field, their deflection velocities interpolated to each particle position, and propagated to the detector. Hits on the detector are binned and Poisson noise is added to produce (C) in \ref{fig:19_sim} and \ref{fig:21_sim}. Then, the flux power spectrum is extracted from (B) and (C) by performing a Fourier transformation, with a cosine tapering applied towards the edges beforehand to preserve periodic boundary conditions. The flux spectra are normalized and a least-squares fit of (\ref{eq:fit}) is performed with $E_{\delta \Psi}(k)$ being the $3.3\,\mathrm{MeV}$ spectra shown in (B) and $\hat{E}_{\delta \Psi}(k)$ the spectra of the $3.3\,\mathrm{MeV}$ predicted data shown in (C). For $n_e=10^{19}\,\mathrm{cm^{-3}}$, the fitted wavenumber cutoff is $k_\mathrm{res}=1270\,\mathrm{cm^{-1}}$, which corresponds to an electron density of $1.05\times10^{19}$; for $n_e=10^{21}$, the fitted wavenumber cutoff is $k_\mathrm{res}=144\,\mathrm{cm^{-1}}$, reproducing an electron density of $9.6\times 10^{20}\,\mathrm{cm^{-3}}$. Finally, to verify the fit, a Gaussian blur with a FWHM width of $a_\mathrm{rms}=4.4M/k_\mathrm{rms}$ (and Poisson noise was applied to the predicted $3.3\,\mathrm{MeV}$ image; comparison of the result shown in (D) with the actual simulated $3.3\,\mathrm{MeV}$ image shows a good reproduction of large-scale features with the correct size. The next step is to validate this diagnostic method on real experimental data.

\subsection{Validation with experimental data}
\label{sec:valmethod:exp}

\subsubsection{Experimental set-up} \label{sec:valmethod:exp:setup}
The experimental campaigns were conducted at the Omega60 laser facility~\cite{Boehly1997}, where previous experiments employing similar setups successfully demonstrated operation of the fluctuation dynamo~\cite{Tzeferacos2018, muller2018, rigby2018, Chen2020, Bott2021b, Bott2022}. This mechanism provides an explanation for the strong magnetic fields observed in astrophysical environments, whereby plasma turbulence exponentially amplifies seed fields by stretching, twisting, and folding magnetic field lines.

\begin{figure}[b]
    \centering
    \includegraphics[width=\linewidth]{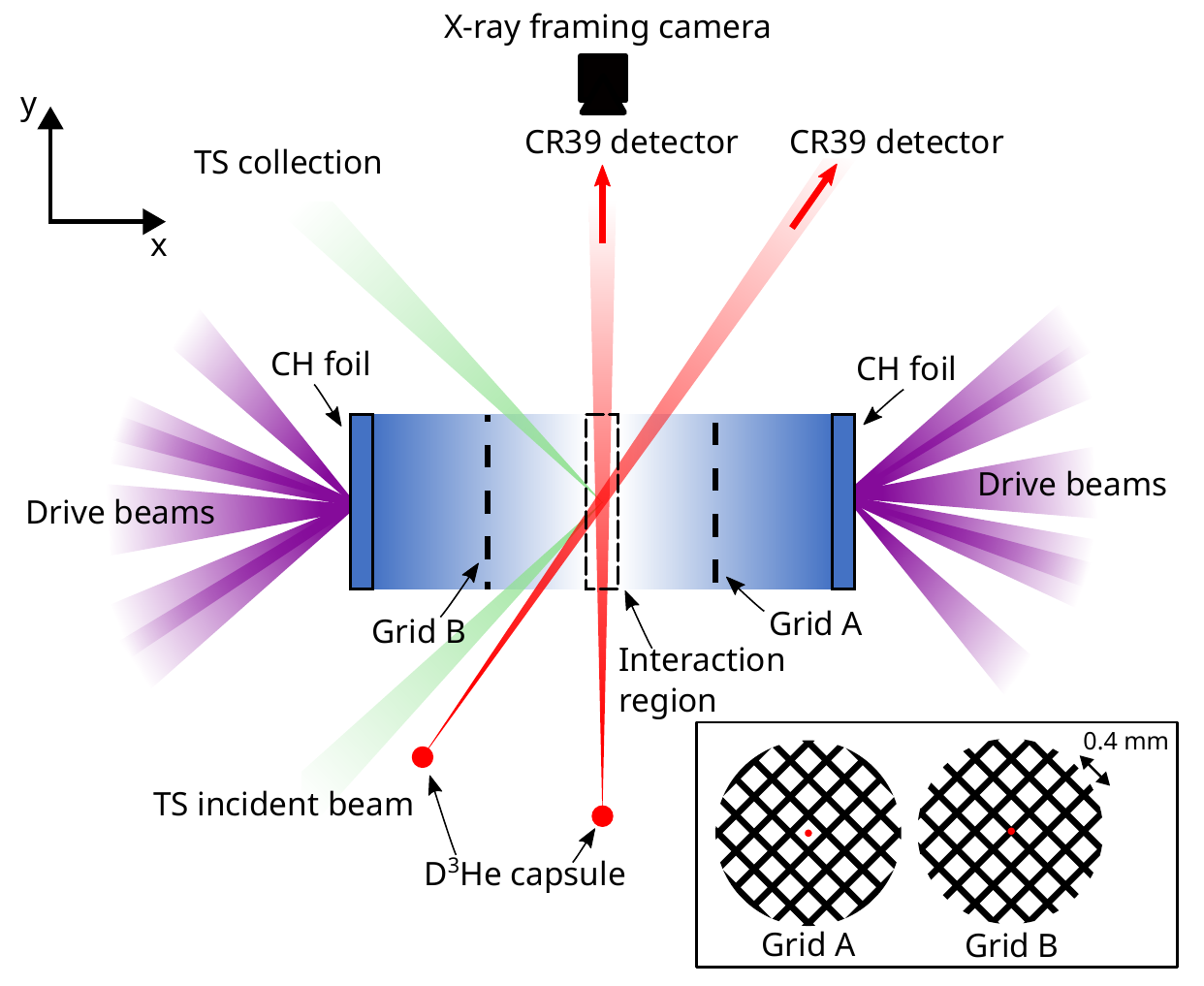}
    \caption{Schematic of the experimental setup. The target foils are composed of CH (equal parts carbon and hydrogen) and are $3\,\mathrm{mm}$ in diameter and $50\,\mathrm{\mu m}$ thick. Each foil is illuminated from the rear by ten $500\,\mathrm{J}$ drive beams, with a $351\,\mathrm{nm}$ wavelength, $800\,\mathrm{\mu m}$ focal spot and a $10\,\mathrm{ns}$ staggered flat pulse. This creates two counterstreaming plasma jets, which pass through grids A and B. The grids are spatially offset such that the holes of one coincide with the rods of the other (bottom right). The jets then collide in the central interaction region, which is probed with Thomson scattering (TS) and proton radiography (the spatial extents of the probes are not to scale). The $526.5\,\mathrm{nm}$ TS probe beam path is shown in green and is not in the plane of the page, probing a cylindrical section with cross-sectional area $50\,\mathrm{\mu m^2}$ and length $1.5\,\mathrm{mm}$ centered on the midpoint of the two foils, with a scattering angle of $63\degree$. The two PRAD lines of sight (not fielded concurrently) are shown in red, with $\theta_\mathrm{p}=90\degree$ and $\theta_\mathrm{p}=55\degree$ relative to the y-axis. The distance between the D\textsuperscript{3}He fusion capsules and the center is not shown to scale, and the CR39 detectors far outside the field of view. X-ray self emission was also collected along the same line of sight as the $\theta_\mathrm{p}=90\degree$ PRAD data with an X-ray framing camera.}
    \label{fig:setup}
\end{figure}

The experimental setup is shown in Fig. \ref{fig:setup}.
Two opposing CH target foils are laser-driven from the rear, producing counterstreaming, supersonic plasma jets. An initial magnetic field is seeded through the Biermann battery mechanism, arising due to the density and temperature gradients formed during laser drive~\cite{Biermann1950}. Grids placed in front of the foils impart a density profile to the jets which are offset with respect to each other. The jets retain this density perturbation and seed magnetic fields before colliding in the central interaction region, where they interpenetrate. This results in shearing motions that drive Kelvin-Helmholtz instabilities, which in turn lead to the turbulent mixing necessary for the fluctuation dynamo to operate.

The strong stochastic fields generated by the dynamo were diagnosed with proton radiography, giving time-resolved measurements of path-integrated fields and the magnetic-energy spectrum along the directions indicated in Fig. \ref{fig:setup}. A D\textsuperscript{3}He fusion capsule is imploded at a distance $r_i=1\,\mathrm{cm}$ away from the center of the plasma, creating a uniformly expanding spherical sheet of $3.3$ and $15.0\,\mathrm{MeV}$ protons. In the first set of experiments, the line connecting the center of the capsule to the geometric center between the two target foils is at an angle $\theta_\mathrm{p}=55 \degree$ to the $x$-axis, while the second had a side-on view of the plasma with $\theta_\mathrm{p}=90\degree$. After traversing the plasma, the protons impinge on the CR-39 nuclear track detector after a distance of $r_s=27\,\mathrm{cm}$, resulting in a magnification of $\mathcal{M}=28$. The detector is specially designed such that the $3.3$ and $15.0\,\mathrm{MeV}$ MeV protons are stopped in separate layers of CR-39.

In addition, self-emitted X-rays were captured by a framing camera along the same line of sight as the $\theta_\mathrm{p}=90\degree$ PRAD (not concurrently) \cite{Bott2021b}. The arrangement consists of a $50\,\mathrm{\mu m}$ pinhole array, $9 \,\mathrm{cm}$ from the interaction region center, and a two-strip microchannel plate (MCP) at a distance of $27\,\mathrm{cm}$, with a filter layer composed of $0.5\,\mathrm{\mu m}$ of polypropene and $150\,\mathrm{nm}$ of aluminum. Electrons exiting the MCP impinge upon a phosphor plate, producing an image that is recorded by a CCD camera. This configuration enables the acquisition of soft X-rays of energy $\gtrsim100 \,\mathrm{eV}$ in a spatially resolved image with a magnification of $\times 2$. Given full ionization, X-ray production is dominated by free-free bremsstrahlung emission. The measured intensity on the CCD is given by $I\propto\int \mathrm{ds}\,n_e^2f(T_e)$, where the integral is performed along the line of sight and $f(T_e)$ encodes the total frequency response of the framing camera and the relative temperature dependence of the measured intensity. As the temperature dependence is weak in the regime of interest, the X-ray framing camera effectively measures electron density.

Thomson scattering with a $30\,\mathrm{J}$, $526.5\,\mathrm{nm}$ laser was performed \cite{Bott2021b}, probing a cylindrical volume with cross-sectional area $50\,\mathrm{\mu m^2}$ and length $1.5\,\mathrm{mm}$ at a scattering angle of $63\degree$. The Thomson scattered signal is spatially resolved and time-integrated over the $1\,\mathrm{ns}$ laser pulse. A modified dynamic form factor which accounts for density fluctuations in the interaction region was fitted to the high-frequency electron plasma wave (EPW) double-peak feature: the position of the peaks is determined by $n_e$.

\subsubsection{Evidence of caustics} \label{sec:valmethod:exp:caustics}
\begin{figure}[h]
    \centering
    \includegraphics[width=\linewidth]{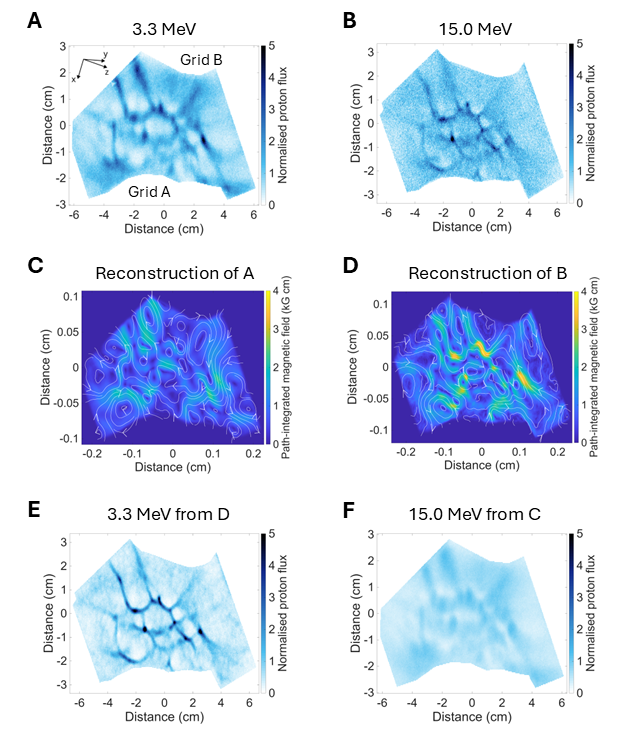}
    \caption{Demonstration of the presence of caustics using field reconstruction and two proton energies. (A) Experimentally measured $3.3\,\mathrm{MeV}$ proton-flux image. (B) same as (A), but with the imaging protons having a higher energy of $15.0\,\mathrm{MeV}$. (C) Reconstructed path-integrated magnetic field obtained by running the PROBLEM code on (A). (D) same as (C), but reconstruction of (B). (E) Predicted $3.3\,\mathrm{MeV}$ proton-flux image, assuming the imaging protons experience the field in (D). For the caustic detection scheme outlined in~\ref{sec:theory:caustics}, this should be compared to (A). (F) Predicted $15.0\,\mathrm{MeV}$ proton-flux image, assuming the imaging protons experience the field in (C), to be compared with (B).}
    \label{fig:prad}
\end{figure}

The presence of caustics in the $3.3\,\mathrm{MeV}$ proton-flux images can be demonstrated using the method described in section~\ref{sec:theory:caustics}. Fig. \ref{fig:prad} shows an illustrative example at $t=30.2\,\mathrm{ns}$, when caustics are likely to be present due to the strong magnetic fields present in the plasma. 
Comparing Fig. \ref{fig:prad}A and E, the morphology and width of the image-flux structures predicted by the reconstructed field in Fig. \ref{fig:prad}D are similar to those in the measured $3.3\,\mathrm{MeV}$ flux image. Conversely, when comparing Fig. \ref{fig:prad}B and F, the two flux images are qualitatively different: the flux structures are much wider in the predicted image. We can conclude that caustics are present in the $3.3\,\mathrm{MeV}$ flux image.

\begin{figure}[h]
    \centering
    \includegraphics[width=\linewidth]{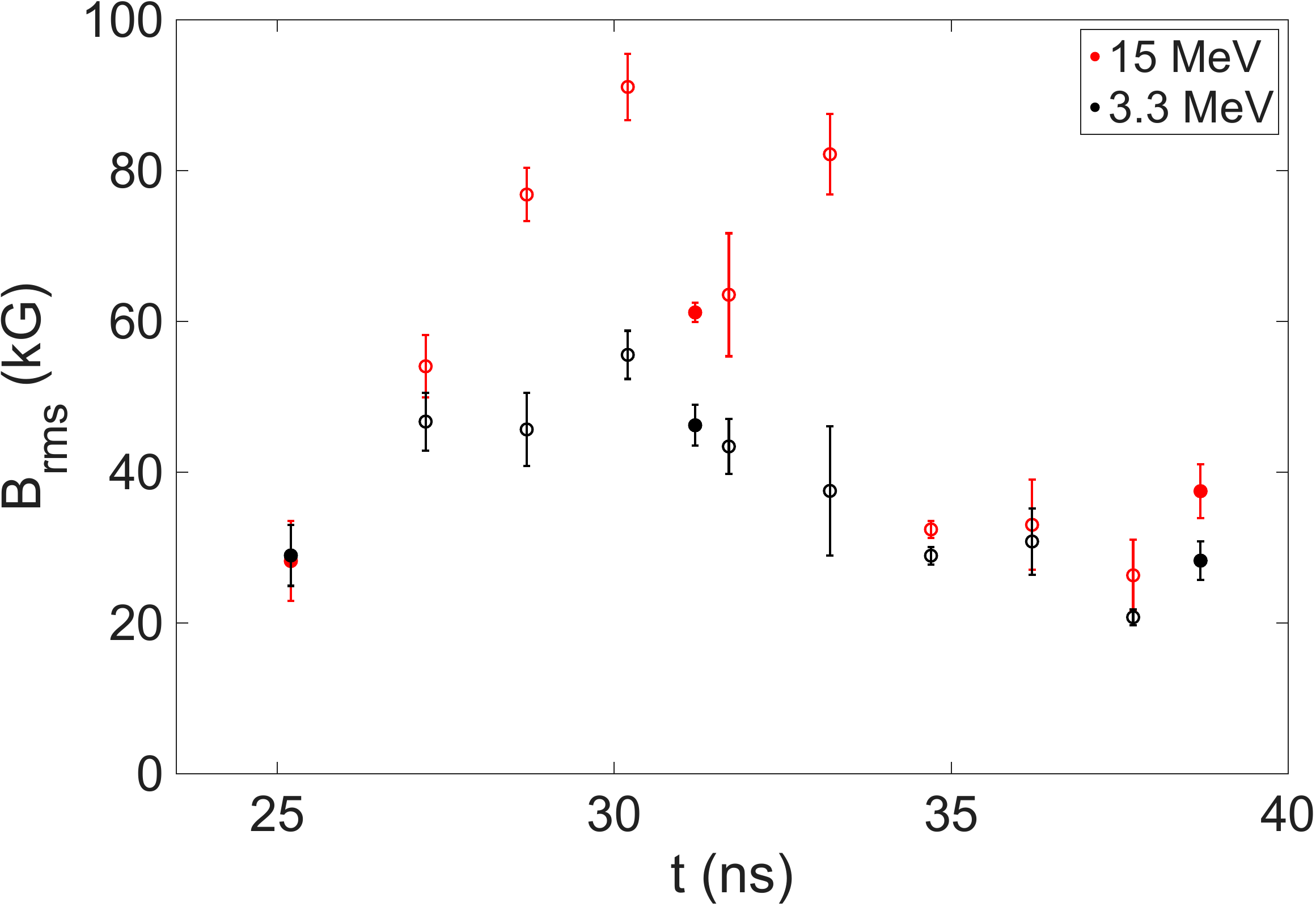}
    \caption{Plot showing the time evolution of the root-mean-squared magnetic field, from reconstruction of $3.3$ and $15.0\,\mathrm{MeV}$ proton-flux images using the PROBLEM code. The values extracted from $15.0\,\mathrm{MeV}$ proton radiography data is shown in red, while those extracted from $3.3\,\mathrm{MeV}$ data are in black; hollow points denote $\theta_\mathrm{p}=55\degree$ data, filled points $\theta_\mathrm{p}=55\degree$. To estimate the uncertainty of these measurements, the interaction region is spatially divided into 3 subregions; the standard deviation of the $B_\mathrm{rms}$ obtained from each subregion is taken to be the error. }
    \label{fig:Bfield}
\end{figure}

The reconstructed fields are significantly weaker for the $3.3\,\mathrm{MeV}$ case because of these caustics (as discussed in II C). This effect can be quantified by comparing the RMS magnetic fields strengths $B_\mathrm{rms}$ for each proton energy. It can be shown that under the assumption of isotropy and homogeneity, $B_\mathrm{rms}$ is directly extractable from the 1D spectrum of the reconstructed path-integrated field~\cite{Bott2017}. As shown in Fig. \ref{fig:Bfield}, the $3.3\,\mathrm{MeV}$ data underpredicts $B_\mathrm{rms}$ for $t=28.7$ to $33.2\,\mathrm{ns}$, which suggests the presence of caustics in the $3.3\,\mathrm{MeV}$ proton-flux images at these times. 

\subsubsection{Comparison with alternative diagnostics of density} \label{sec:valmethod:exp:data}
\begin{figure}
    \centering
    \includegraphics[width=\linewidth]{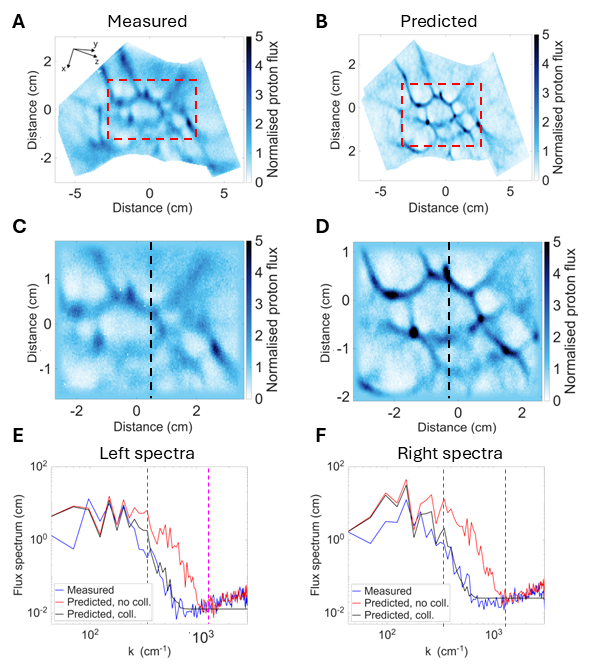}
    \caption{Demonstration of the process of extracting a density measurement from PRAD data. (A) shows the measured $3.3\,\mathrm{MeV}$ proton-flux image at $t=30.2\,\mathrm{ns}$ (as shown in Fig. 2A), with the rectangular region over which analysis is to be performed indicated in red. (B) shows a $3.3\,\mathrm{MeV}$ `predicted' flux image, generated using the same approach as described in section \ref{sec:valmethod:sims}, with the selected region indicated in red. (C) and (D) show the cropped rectangular regions from (A) and (B), respectively, with a cosine tapering applied. The vertical dashed black line indicates the axis along which the regions are divided in half and reflected. (E) and (F) show log-log plots of flux spectra obtained from the left and right halves of the rectangular regions, respectively. The flux spectrum of the measured $3.3\,\mathrm{MeV}$ proton-flux image is shown in blue, the flux spectrum from the predicted flux image is shown in red with a normalization applied such that the mean fluence is equal to the measured case, and the fitted flux spectrum, using (\ref{eq:fit}), is shown as the dashed black line. The vertical dashed lines in magenta bookend the range of $k$ over which the least-squares fit is performed. }
    \label{fig:densityprocess}
\end{figure}

For this analysis, we follow a similar treatment to that performed on the synthetic data in section~\ref{sec:valmethod:sims}. We assume the reconstructed deflection field from the $15.0\,\mathrm{MeV}$ data is the `true' field that the $3.3\,\mathrm{MeV}$ imaging protons also see. Therefore, by generating a number of test particles ($3.3\,\mathrm{MeV}$ protons) which follow the initial fluence profile, and propagating them through this field, we arrive at a synthetic proton-flux image without any caustic structures. In addition, a Gaussian blur with a width related to the finite source size is applied~\cite{Manuel2012}, as well as Poisson noise. Fig. \ref{fig:densityprocess} shows an example of the procedure used to extract a density measurement from the measured and synthetic proton-flux images at $t=30.2\,\mathrm{ns}$.  

First, the flux images are cropped to a rectangular region of interest. A cosine tapering with a width ${\sim}10\%$ of the size of the region is applied such that the flux intensity tends towards zero as you move towards the edges. This is an important step as it ensures the image satisfies periodic boundary conditions: the sharp edges in the unprocessed data can cause spectral leakage and affect the characteristic caustic spectrum in the high wavenumber range. Then, the flux images are spatially divided into two halves and reflected across the same axis to preserve periodic boundary conditions. The drawback of performing these pre-processing steps is a reduction in resolution in k-space; more quantitatively, these steps introduce a spectral broadening that is numerically estimated to have a width of ${\sim}25\,\mathrm{cm^{-1}}$. This is on the order of, but smaller than, the $k$-space resolution, so the pre-processing is not expected to significantly change the characteristic spectrum. Next, we perform Fourier analysis on each half of the measured and predicted $3.3\,\mathrm{MeV}$ proton-flux images to obtain the 1D flux power spectra. A least-squares fit of (\ref{eq:fit}) is performed on each half, with $E_{\delta \Psi}(k)$ defined as the 1D flux spectrum of the $3.3\,\mathrm{MeV}$ measured proton-flux image, and $\hat{E}_{\delta \Psi}(k)$ the flux spectrum of the flux image predicted from the $15.0\,\mathrm{MeV}$ reconstructed fields. This fit is performed over a specific range in $k$, starting from where the measured flux spectrum begins to steepen and diverge from the true spectrum to the $k$ at which the two converge again, as shown in Fig. 4E and F. This prevents random variations in the low $k$ range from dominating the fit; the high wavenumber tails of the measured and predicted spectra can also differ substantially if noise representative of the measured flux image is not artificially applied to the predicted flux image correctly. Finally, $n_e$ in each half is recovered from the fitted $k_\mathrm{res}$ using (\ref{eq:rms_deflection}) to (\ref{eq:fit}) and averaged. Because the plasma is approximately fully ionized with equal parts carbon and hydrogen, it follows that $\nu_{\perp}^{p\vert i} = \nu_{\perp}^{p\vert C}+\nu_{\perp}^{p\vert H}$, with $n_C=n_H=n_e/7$. Following the discussion in~\ref{sec:theory:exp_lim}, for this experimental setup with a source size of $40\,\mathrm{\mu m}$, the minimum measurable density is $\sim 10^{16}\,\mathrm{cm^{-3}}$. In principle, the technique could be applied on a grid for a spatially resolved measurement, but the characteristic spectrum of caustics requires a sufficiently high number of caustics to be sampled. For a flux image containing many caustics with clearly defined positions, an irregular density map local to the regions with caustics could be constructed.     

\begin{figure}
    \centering
    \includegraphics[width=\linewidth]{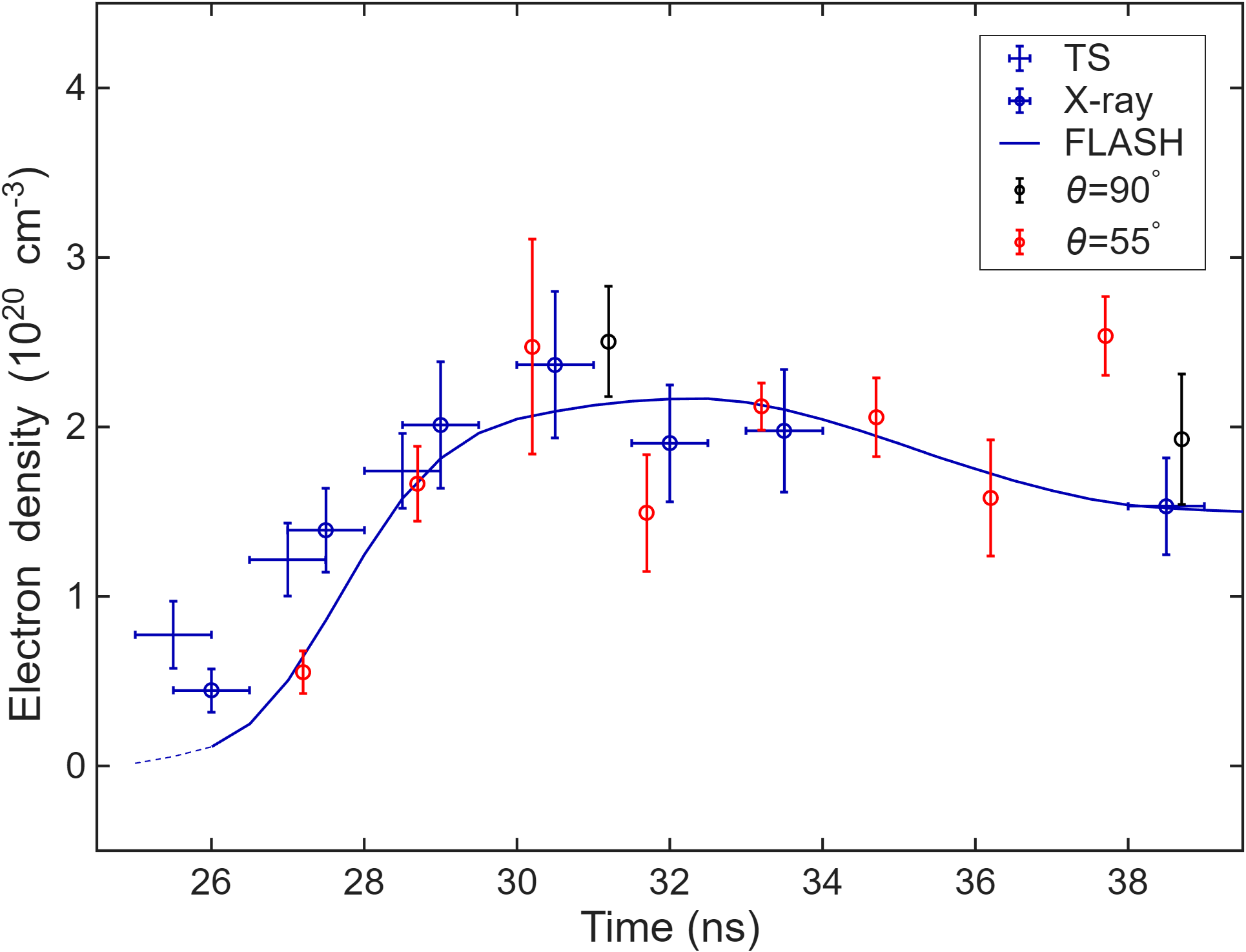}
    \caption{Plot of free electron density at different times. Values extracted from experimental PRAD data are shown in red for the angled ($\theta=55\degree$) line-of-sight and black for the side-on ($\theta=90\degree$) configuration. The uncertainties of the densities extracted from PRAD are calculated by taking the standard deviation of the values obtained from each half-image. The density measured by Thomson scattering, x-ray imaging and FLASH MHD simulations are shown in blue.}
    \label{fig:densities}
\end{figure}

As collisional scattering accumulates over the interaction path, knowledge of the path length $l_n$ is required for a volumetric measurement of $n_e$. For $\theta_\mathrm{p}=90\degree$, the extent of the interaction region sampled by the imaging protons is clearly defined as $l_n\simeq 3\,\mathrm{mm}$; this enables the calculation of the path-averaged density $n_e$. As shown in Fig. 7, the $n_e$ extracted from the $\theta_\mathrm{p}=90\degree$ PRAD data shows good agreement with self-emission X-ray imaging, Thomson scattering, and FLASH MHD simulations. 

\begin{figure}
    \centering
    \includegraphics[width=\linewidth]{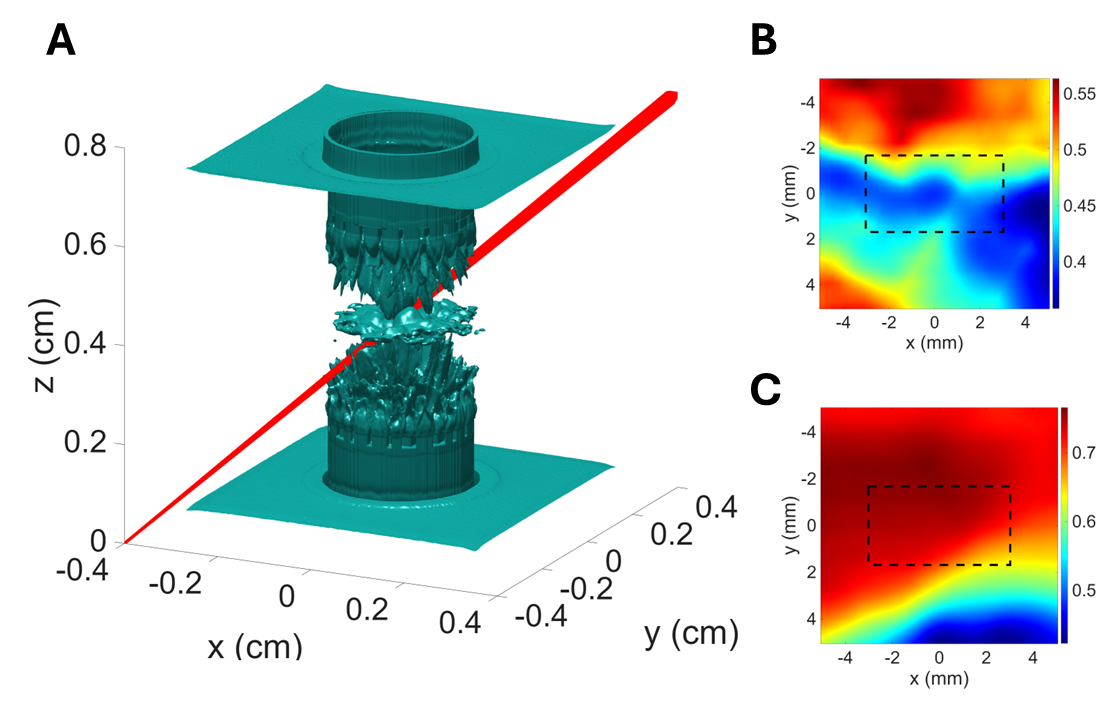}
    \caption{Analysis of FLASH magnetohydrodynamic simulation data for estimating the contribution of density outside of the interaction region towards collisional broadening. (A) shows an isosurface with a density of $n_e=10^{20}\,\mathrm{cm^{-3}}$, showing the structure of the inflow jets and the formation of an interaction region at $t=32\,\mathrm{ns}$ after the start of laser drive. The sampled proton trajectories are shown as the red beam and can be seen to pass through both the inflow jets and the interaction region. (B) shows a map of the contribution from the interaction region to the total collisional scattering angle at the detector, $t=32\,\mathrm{ns}$ after the start of laser drive; (C) shows the ratio at $t=37\,\mathrm{ns}$. The black dashed rectangle demarcates the approximate extent of the interaction region and the correction factor is obtained by averaging over this region.}
    \label{fig:pathsim}
\end{figure}

In the case of $\theta_\mathrm{p}=55\degree$, although the time evolution of $l_n$ is a known quantity, the geometry of the setup means that the imaging protons also traverse the inflow jets before reaching the detector. This adds a non-negligible contribution to the collisional scattering and must be corrected in order to obtain an accurate measurement of $n_e$ in the interaction region. Due to spatial variations in density, temperature and ionization state within the inflow jets, determining this correction is not a straightforward process. To this end, we employ FLASH MHD simulations, which provide three-dimensional, spatially resolved measurements of the relevant plasma parameters. A point source and a detector are placed at their respective physical locations, and a large number of possible paths are generated within the cone of the solid angle subtended by the detector, as shown in Fig. \ref{fig:pathsim}(A). For each generated path, the plasma parameters are sampled along the path, and the contribution to the collisional scattering angle $\theta_C$ is summed; the ratio of the contribution from within the interaction region to the total $\theta_C$ is calculated over the detector area, producing Fig. \ref{fig:pathsim} (B) and (C). A correction factor is obtained by averaging over the spatial extent of the interaction region, for FLASH data from 6 timesteps from $t=24.5$ to $37\,\mathrm{ns}$ after the start of laser drive; linear interpolation is performed to find the correction factors at the times experimental PRAD data was collected.

The $n_e$ extracted from the $\theta_\mathrm{p}=55\degree$ PRAD data shows good agreement with other diagnostics following the application of the correction factor.  This is especially true for $t=30.2$ to $33.2\,\mathrm{ns}$, where $B_\mathrm{rms}$ is highest and caustic features are most prominent. On the other hand, there is an overestimation of density at $t=37.7\,\mathrm{ns}$ for the $\theta_\mathrm{p}=55 \degree$ data. This could possibly be attributed to the fact that caustic structures are weaker at this time, as evidenced by Fig. \ref{fig:Bfield} and the analysis outlined in III C. However, the $\theta_\mathrm{p}=90\degree$ data recovers only a slightly higher density within the uncertainty bounds, suggesting that the deviation could also be affected by the correction factor for the inflow jet contribution. This necessary procedure highlights the importance of a clear line-of-sight for the imaging protons, to preserve the accuracy of the $n_e$ measurement and minimize reliance on simulations in interpreting results. 

\section{Conclusion} \label{sec:conclusion}
In this paper, we have presented a technique for simultaneous diagnosis of electromagnetic fields and the electron number density $n_e$ in laser plasmas. We build on the existing diagnostic of proton radiography, commonly used to diagnose electromagnetic fields in laser-plasma experiments by passing imaging protons through the plasma and analyzing the image-flux distribution measured at a detector. As a secondary process, the imaging protons also undergo collisional scattering, which is a density-dependent effect. In previous experiments, this effect has only been leveraged in higher-density laser plasmas ($n_e \gtrsim 10^{22} \, \mathrm{cm}^{-3}$ to measure density. 
In this work, we have outlined a procedure for determining the electron number density from proton radiography at significantly lower densities. The technique only requires at least two species of imaging protons with different energies: a lower energy where caustic structures due to electromagnetic fields are present in the proton-flux image, and a higher energy where the proton-flux image is in the linear or nonlinear injective contrast regime. This condition of multiple proton energies can be met by conventional proton sources, including those generated by high-intensity lasers and fusion implosion capsules, and no other additional modifications are necessary. Results show good agreement with other well-established density diagnostics including Thomson scattering and x-ray imaging. 

This research has implications for a broad range of laser-plasma experiments, with particular applications to laboratory astrophysics. Proton imaging is routinely used to diagnose electromagnetic fields in experiments investigating magnetic reconnection, turbulence, jets, shocks, and plasmas instabilities, for which the plasma of interest has an electron number density in the range $n_e \sim 10^{18}$-$10^{21} \; \mathrm{cm}^{-3}$~\cite{Schaeffer2023}. The technique reported here opens up the possibility of obtaining simultaneous measurements of path-integrated electromagnetic fields and areal electron densities along the same line of sight for this range of densities. Such measurements could prove to be valuable where the plasma dynamics depend on correlations between density structure and electromagnetic fields, whilst also avoiding uncertainties associated with comparing measurements carried out along different lines of sight or on different experimental shots. More generally, the technique provides an additional way to constrain the plasma density without requiring a dedicated density diagnostic, and could therefore prove especially useful at facilities where more established diagnostics such as interferometry, Thomson scattering, or x-ray imaging are unavailable or impractical because of geometric limitations or other experimental constraints.  

\section*{Data Availability}
The data that support the findings of this article are publicly available [1]. The FLASH code, used for simulations reported in this paper, is publicly available and under licence. Access to the FLASH code can be requested through the Flash Center website. The version of the code associated with the paper is maintained under version control. Equation of state and opacity data used in the simulations are proprietary, generated with Prism Inc. software PROPACEOS.

\section*{Acknowledgements}
A.F.A.B. was supported for this research by funding from the UKRI (grant number MR/W006723/1). The work of G.G. was supported in part by the UK Research and Innovation (UKRI) Frontier Research Guarantee under Grant No EP/Y035038/1 and by Science and Technology Facilities Council (STFC) under Grant No ST/W000903/1. We acknowledge support by the U.S. DOE NNSA under Awards DE-NA0002724, DE-NA0003605, DE-NA0003934, DE-NA0004144, DE-NA0004147, and Subcontracts 630138 and C4574 with LANL; the U.S. DOE Office of Science under Award DE-SC0021990; and the NSF under Awards PHY-2033925 and PHY-2308844. The software used in this work was developed in part by the U.S. DOE NNSA- and U.S. DOE Office of Science-supported Flash Center for Computational Science at the University of Chicago and the University of Rochester.

\newpage

\bibliography{apssamp}

\end{document}